\documentstyle[aaspp4]{article}

\newcommand{\gb}{\gamma_{\rm j}\beta_{\rm j}}

\begin{document}
\title{The Nuclear Jet in M81}
\author{Heino Falcke}
\affil{Astronomy Department, University of Maryland, College Park,
MD 20742-2421 (hfalcke@astro.umd.edu)}

\begin{abstract}
In this paper we apply the jet-disk symbiosis model developed for Sgr
A* to M81* -- the nucleus of the nearby galaxy M81. The model
accurately predicts radio flux and size of M81* for the observed
bolometric luminosity of the nuclear source with no major free
parameter except for the inclination angle. We point out that the
usually applied free, conical jet emission model implies a
longitudinal pressure gradient that must lead to a moderate
acceleration of the jet along its flow direction. This, usually
neglected, gradual acceleration naturally accounts for the inverted
spectrum and the size/frequency relation of M81* and may be a general
feature of radio cores.  M81* is so far the best case for a radio-loud
jet nature of the compact radio core in the nucleus of a nearby spiral
galaxy. The fact that one can account for Sgr A* and M81* with the
same model by simply changing the accretion rate, strongly supports
the jet-disk symbiosis model as an explanation for the compact radio
cores of galaxies in general.
\end{abstract}

\keywords{Galaxies: active, indivdual:
M81---Galaxies: jets---accretion, accretion disks---black hole
physics---Galaxy: center}

\section{Introduction}
Quite a few nearby galaxies seem to have compact radio cores in their
nuclei, prominent cases are the Milky Way (Sgr A*), M31 and M81. Those
radio cores resemble the cores of radio loud quasars, showing a very
high brightness temperature and a flat to inverted radio spectrum that
extends up to submm wavelengths. Several models have been developed to
explain those cores in the context of black hole accretion: Melia
(1992a\&b) suggested a spherical accretion model for Sgr A* and, what
he called, M31*. Falcke et al. (1993) and Falcke, Mannheim, \&
Biermann (1993, hereinafter FMB93) proposed an alternative model,
where Sgr A* was explained as the core of a radio jet, fed by an
underluminous, starving accretion disk (see also Falcke 1996a, for a
review, and Falcke \& Heinrich 1994 for M31*).  This model evolved
into the jet-disk symbiosis approach (Falcke \& Biermann 1995,
hereinafter FB95; Falcke 1996b) that unified the explanation for radio
cores of quasars (Falcke, Malkan, \& Biermann 1995, hereinafter
FMB95), galactic jet sources (Falcke \& Biermann 1996) and Sgr A* into
a single picture. The basic idea was to postulate that black
holes, jets and disks form closely coupled systems with little
variations from one system to another except for the accretion rate.

Recent VLBI (Bietenholz et al. 1996), multiwavelengths (Ho, Filipenko
\& Sargent 1996) and submm observations (Reuter \& Duschl 1996) have
now shown that the radio core in M81 -- which in analogy to Sgr A* and
M31*, we will label M81* hereafter -- is very similar to Sgr A*. 
This wealth of data now makes M81* an excellent laboratory to test the
jet-disk symbiosis in detail.

\section{Free Jet with Pressure Gradient}
One of the interesting findings of the VLBI observations (de Bruyn et
al. 1976; Bartel et al. 1982; Bietenholz et al. 1996) is that the
radio core of M81* is elongated and its size is intrinsically
frequency dependent with $r\propto\nu^{-0.8\pm0.05}$. A frequency
dependent size was one of the basic predictions of the jet model and
excludes any homogenous, optically thin models (e.g. as in Duschl \&
Lesch 1994). However, the measured size-frequency relation is slightly
shallower than in the Blandford \& K\"onigl (1979, hereinafter BK79)
free jet model adopted by FB95, and the radio spectrum of Sgr A* and
M81* are both slightly inverted rather than flat as in the simple
model. Here we show, that a self-consistent treatment of the BK79
model implies a weak acceleration of the bulk jet flow due to its
longitudinal pressure gradient which can naturally explain the
observed size- and spectral indices in compact radio cores.

\subsection{Longitudinal Velocity Profile}
As in FB95, we shall describe the jet core within the framework of
relativistic gasdynamics of a relativistic gas with adiabatic index
$\Gamma=4/3$. We consider only the supersonic regime and impose as
boundary condition, that the jet expands freely with its initial
soundspeed $\beta_{\rm s}$ without any lateral gradients behind the
jet nozzle. This leads to the familiar conical jet with
$B^2/4\pi\propto r^{-2}$ (BK79), where adiabatic losses due to lateral
expansion need not be considered. For a confined jet scenario
(e.g. Sanders 1983) at larger scales the adiabatic losses could,
however, be quite severe with energy densities scaling as $r^{-2\Gamma}$.

Nonetheless, even in a simple, free jet at least some work due to
expansion will be done, because there is always a longitudinal
pressure gradient, which will accelerate the jet along
its axis. This acceleration is described by the
z-component of the modified, relativistic Euler equation
(e.g. Pomraning 1973, Eq.~9.171) in cylindrical coordinates where we
set $\partial/\partial r=0$ and $\partial/\partial \theta=0$
\begin{equation}\label{euler1}
\gb {\partial\over\partial z}\left(\gb{\omega\over n}\right)=-{\partial\over\partial z}P.
\end{equation}
Here, $\omega=m_{\rm p}nc^2+U_{\rm j}+P_{\rm j}$ ist the enthalpy
density of the jet, $U_{\rm j}$ is the internal energy density, $n$ is
the particle density, and $P_{\rm j}=(\Gamma-1)U_{\rm j}$ is the
pressure in the jet (all in the local rest frame). For the `maximal
jets' -- the radiatively most efficient type of jet (FB95) -- to be
discussed here, we demand the equivalence of internal energy and
kinetic (or better rest mass) energy $U_{\rm j}\simeq m_{\rm p}nc^2$,
hence $\omega=(1+\Gamma)U_{\rm j}$ and ${\omega/n}=(1+\Gamma)m_{\rm p}
c^2=$const at the sonic point $z=z_{0}$.  In the free jet with conical
shape the energy density evolves as $U_{\rm j}\propto
\left(\gb\right)^{-\Gamma}z^{-2}$ and we do not consider any loss
mechanisms other than adiabatic losses due to the longitudinal
expansion. Using the relations mentioned above, the Euler equation
becomes

\begin{equation}\label{euler2}\label{v}
{\partial\gb\over\partial z}
\left({\left({\Gamma+\xi\over\Gamma-1}\right)(\gb)^2-\Gamma\over\gb}\right)={2\over z}
\end{equation}
with $\xi=\left(\gb/(\Gamma(\Gamma-1)/(\Gamma+1))\right)^{1-\Gamma}$.
For $\xi\sim1$ this is analogous to the Euler equation for the well
known isothermal, pressure driven winds, with the wind speed replaced
by the proper jet speed and the sound speed is constant and fixed by
the required equivalence between internal and rest mass energy density
at a value $\beta_{\rm s}=\sqrt{(\Gamma-1)/(\Gamma+1)}\sim0.4$.

The asymptotic solution of Eq.~\ref{euler2} for $\xi=$const and
$z\rightarrow\infty$ is $\gb \propto 2\sqrt{\ln z}$. If we ignore
terms of the order $\ln \gb$ we could approximate the solution by $\gb
\simeq \sqrt{{\Gamma-1\over\Gamma+1}\left(\Gamma+4\ln
\left(z/z_{0}\right)\right)}$. For the following calculations we will,
however, use the exact, numerical solution to Eq.~\ref{euler2}
($\xi\neq$const), but the deviation from the approximate, asymptotic
solution is rather small.

\subsection{Plasma Properties}
The basic ideas how to derive the synchrotron emissivity and basic
properties of a jet in a coupled jet-disk system have been described
extensively in FMB93 and FB95. The power $Q_{\rm j}=q_{\rm j}\dot
M_{\rm disk} c^2 = q_{\rm j/l} L_{\rm disk}$ of {\em one} jet cone is
a fixed fraction of the disk luminosity, relativistic particles and
magnetic field $B$ are in equipartition within a factor $k_{\rm e+p}$
-- which we hereafter set to one, and the energy fluxes are conserved
along the conical jet (for a moment we will forget the adiabatic
losses). Mass conservation requires that the mass loss in the jet
$\dot M_{\rm jet}$ is smaller than the mass accretion rate in the disk
$\dot M_{\rm disk}$, thus $q_{\rm m}=\dot M_{\rm jet}/\dot M_{\rm
disk} < 1$, while mass and energy of the jet are coupled by the
relativistic Bernoulli equation $\gamma_{\rm j} q_{\rm
m}\left(1+\beta_{\rm s}^2/(\Gamma-1)\right)=q_{\rm j}$ (FMB93).

Here we will make use of the same logic and notation but with three changes
with respect to FB95, namely (1) applying the velocity law Eq.~\ref{v},
(2) neglecting the energy contents in turbulence and (3) considering
only a quasi monoenergetic energy distribution for the electrons at a
Lorentz factor $\gamma_{\rm e}$ (usually $\ga100$). The latter is
indicated by the steep submm-IR cut-off of Sgr A* (Zylka et al. 1995)
-- and probably also M81* (Reuter \& Lesch 1996) -- which is different
from typical Blazar spectra and precludes an initial electron powerlaw
distribution (as in FMB93).

The semi-opening angle of the jet is $\phi=\arcsin\left(\gamma_{\rm
s}\beta_{\rm s}/\gb\right)$, and the magnetic field in the comoving
frame of a maximal jet with the given sound speed, $L_{\rm
disk}=L_{41.5}10^{41.5}$ erg/sec, $q_{\rm j/l}=0.5 q_{0.5}$, and
$z_{16}=z/10^{16}$cm becomes (see Eq.~19 in FB95)
\begin{equation}\label{b}
B=0.6\,{\rm G}\; \sqrt{\beta_{\rm j} L_{41.5} q_{0.5}}z_{16}^{-1}.
\end{equation}

The number of relativistic electrons that are to be in equipartition
with the magnetic field are a fraction $x_{\rm e}=n_{\rm e}/n$ of the
total particle number density, and the energy density ratio between
relativistic protons and electrons is $(\mu_{\rm p/e}-1)$. From the
energy equation $\gamma_{\rm j}\omega\gb c\pi r^2=q_{\rm j/l}L_{\rm
disk}$ we find that the characteristic Lorentz factor and electron
density required to achieve equipartition are
\begin{eqnarray}\label{gamma}
\gamma_{\rm e}&=&m_{\rm }/\left(4\Gamma m_{\rm e}\mu_{\rm p/e}x_{\rm e}\right)=344/\left(\mu_{\rm p/e}x_{\rm e}\right)\\
n_{\rm e}&=&45\,{\rm cm^{-3}}\;\beta_{\rm j} L_{41.5} q_{0.5} x_{\rm
e}z_{16}^{-2}\label{n}
\end{eqnarray}
Finally, to incorporate adiabatic losses we now have to make the
following transitions with respect to the equations in FB94
\begin{eqnarray}\label{adloss}
&&\gb\rightarrow\gb(z),\;\;\gamma_{\rm e}\rightarrow\gamma_{\rm e}\cdot(\gb(z)/\gb(z_{\rm 0}))^{1/3}\nonumber\\
&&B\rightarrow B\cdot(\gb(z)/\gb(z_{\rm 0}))^{1/6}.
\end{eqnarray} 

\subsection{Synchrotron Emission}
The local spectrum of the jet will be $F_{\nu}\propto\nu^{1/3}$
between the synchrotron self-absorption frequency $\nu_{\rm ssa}(z)$
and the characteristic frequency $\nu_{\rm c}(z)$ = $3 e \gamma_{\rm
e}^2 \sin\alpha_{\rm e} B(z) / 4 \pi m_{\rm e}c$ (here we will use
an average pitch angle $\alpha_{\rm e}=60^\circ$). For the maximal
jet using Equations \ref{b}, \ref{gamma}, \ref{n} and
\ref{adloss}, we have in the observers frame
\begin{equation}\label{nuc}
\nu_{\rm c}(z)=100\,{\rm GHz}\; {\cal D} {\sqrt{L_{41.5}
q_{0.5}}}/\left(
\sqrt{\beta_{\rm j}}\gamma_{\rm j} x_{\rm e}^2 \mu_{\rm p/e}
z_{16}\right)
\end{equation}
with the Doppler factor ${\cal D}=1/\gamma_{\rm j}\left(1-\beta_{\rm
j} \cos i\right)$ for an inclination $i$ of the jet axis to the line
of sight.

To find the local synchrotron self-absorption frequency $\nu_{\rm
ssa}$ we have to solve the equation $\tau=n_{\rm e} \sigma_{\rm
sync}(\nu) \cdot 2 r_{\rm j}/\sin i=1$ and transform into the
observers frame. The jet radius is $r_{\rm j}\sim \phi z$ and the
synchrotron self-absorption cross section for a monoenergetic electron
distribution is given by $\sigma_{\rm sync}=$ $4.9\cdot10^{-13}\,{\rm
cm}^2$ $(B/{\rm G})^{2/3}\gamma_{\rm e}^{-5/3}(\nu/{\rm GHz})^{-5/3}$,
yielding

\begin{equation}\label{nus}
\nu_{\rm ssa}=3.3\,{\rm GHz}\;{\cal D}{{\beta_{\rm
j}^{1/5}L_{41.5}^{4/5}q_{0.5}^{4/5}x_{\rm e}^{8/5}\mu_{\rm p/e}}\over
\gamma_{\rm j}^{3/5} ({\cal D}\sin i)^{3/5} z_{16}}.
\end{equation}

If we now use the velocity profile Eq.~\ref{v} we can invert
Eq.~\ref{nuc} to find the size of the jet as a function of observed
frequency. For the asymptotic regime $z\gg z_{\rm 0}$,  $\nu_{\rm c}$ is
approximated for a fixed inclination angle to within a few per-cent by
a powerlaw in $z$, such that

\begin{eqnarray}\label{size}
z_{\rm c}&=(z_{{\rm c,0}}/\sin i)(\mu_{\rm p/e}x_{\rm e})^{-2\xi} \left(q_{0.5}L_{41.5}\right)^{\xi/2}z_{13.7}^{1-\xi}\nu_{10.3}^{-\xi}
\end{eqnarray}
where the parameters are $z_{13.7}=z_0/3$~AU, $\nu_{10.3}=\nu/22$~GHz,
$\xi=(0.99,0.95,0.9,0.89,0.88)$ and $z_{{\rm
c,0}}=(500,1200,1100,900,700)$ AU for $i=(5^\circ,20^\circ, 40^\circ,
60^\circ, 80^\circ)$.

The total spectrum of the jet is obtained by integrating the
synchrotron emissivity $\epsilon_{\rm sync}$ along the jet:
$F_\nu=(4\pi D^2)^{-1}\int_{z({\nu_{\rm ssa}})}^{z({\nu_{\rm c}})}
\epsilon_{\rm sync}\pi r^2 {\rm d}z$, which can again be 
approximated using powerlaws:

\begin{eqnarray}\label{flux}
F_{\nu}&=745\,{\rm mJy}(q_{0.5}L_{41.5})^{1.46}\mu_{\rm
p/e}^{.17}x_{\rm e}^{1.17}z_{13.7}^{0.08}\nu_{10.3}^{0.08}\nonumber\\
&-337\,{\rm mJy}(q_{0.5}L_{41.5})^{1.54}\mu_{\rm p/e}^{.92}x_{\rm
e}^{2.1}z_{13.7}^{0.08}\nu_{10.3}^{0.08},\\
F_{\nu}&=247\,{\rm mJy}(q_{0.5}L_{41.5})^{1.42}\mu_{\rm
p/e}^{.33}x_{\rm e}^{1.33}z_{13.7}^{0.16}\nu_{10.3}^{0.16}\nonumber\\
&-143\,{\rm mJy}(q_{0.5}L_{41.5})^{1.48}\mu_{\rm p/e}^{.85}x_{\rm
e}^{1.95}z_{13.7}^{0.15}\nu_{10.3}^{0.15},\\
F_{\nu}&=106\,{\rm mJy}(q_{0.5}L_{41.5})^{1.40}\mu_{\rm
p/e}^{.40}x_{\rm e}^{1.40}z_{13.7}^{0.20}\nu_{10.3}^{0.20}\nonumber\\
&-71\,{\rm mJy}(q_{0.5}L_{41.5})^{1.45}\mu_{\rm p/e}^{.81}x_{\rm
e}^{1.89}z_{13.7}^{0.19}\nu_{10.3}^{0.19},
\end{eqnarray}
for $D=3.25$ Mpc, and $i=(20^\circ,40^\circ,60^\circ$)
respectively. This means that the resulting spectrum is no longer flat
but tends to be inverted with $\alpha\sim0.15-0.27$ for
$i=30^\circ-90^\circ$. The spectral index, like the size index, is a
function of the inclination angle of the jet and tends to become
flatter for smaller $i$.

\section{Application to M81*}
We can now apply this formalism to the situation in M81*, where we have
five observed quantities, namely the time averaged spectral index
$\alpha\sim0.2\pm0.2$ (Reuter \& Lesch 1996), the flux $F_\nu(22{\rm
GHz})\simeq120$ mJy, the size $z(22{\rm GHz})\simeq550$ AU, the
frequency dependence of the size $z\propto\nu^m$ with $m={-0.8\pm0.05}$
(Bietenholz et al. 1996), and the bolometric luminosity of the nuclear
source (Ho, Filipenko, \& Sargent 1996) which is $L_{\rm disk}\sim10^{41.5}$
erg/sec.

Most of the model parameters are already well constrained. As the
cosmic-ray ratio $\mu_{\rm p/e}$ and the fractional number of
electrons $x_{\rm e}$ are almost interchangeable we keep the latter
fixed at $x_{\rm e}=1$. The jet-disk ratio $q_{\rm j/l}$ was
determined in FMB95 to be $\sim0.15$ for a flat spectrum. To bring
this model with an inverted spectrum and a peak at submm wavelengths
to the same fluxscale one has to increase this value to $q_{\rm
j/l}=0.5$ where disk luminosity and jet have equal powers; here we do
not intent to change this value. The size of the nozzle $z_0$ enters
only weakly into the equations; if we assume that the high-frequency
cut-off which is somewhere around 1000 GHz correspond to the size
scale of the nozzle (in analogy to Sgr A* in Falcke 1996) we find by
extrapolation that $z_0\sim3$ AU as an order of magnitude
estimate. For a nozzle size of 10-100 gravitational radii this
correponds to black hole masses of $0.3-3\cdot10^7M_\odot$ which is
consistent with dynamical estimates (Ho et al. 1996).

This leaves us with two main input parameters: $i$ and $\mu_{\rm
p/e}$.  The cosmic-ray ratio determines the electron Lorentz factor
and is well constrained; by definition $\mu_{\rm p/e}\ge1$ and from
the condition $\nu_{\rm ssa}<\nu_{\rm c}$ we here have $\mu_{\rm
p/e}\la3x_{\rm e}^{-6/5}$ (Eqs.~\ref{nuc}\&\ref{nus}). Figure 1 shows
the predicted sizes and fluxes of the model for various $\mu_{\rm
p/e}$ and $i$, indicating that in M81* $i\sim30-40^\circ$ and
$\mu_{\rm p/e}\sim1.5$ (FMB95 used $\mu_{\rm p/e}=2$ for quasars). The
predicted spectral index then is $\alpha\sim0.15-0.19$ and the
predicted size index is $-m=0.9\;-\; 0.92$. Considering the simplicity
of the model, those values are reasonably close to the observed ones.

\section{Discussion}
The jet-disk symbiosis emission model by FMB93 and FB95 which was
initially developed to explain the radio core of Sgr A* and radio loud
quasars can also explain the nuclear radio source in M81 in detail.
Simply by scaling the accretion disk luminosity by several powers of
ten to the observed value properly predicts flux, size, spectral index, and
size index of the VLBI radio source.

If one inserts M81* into the universal $L_{\rm disk}/$radio
correlation presented in Falcke \& Biermann (1996) it falls right onto
the line that connects Sgr A* and radio loud quasars\footnote{This was
already shown in a preliminary way in Falcke (1994, Fig. 8.1).}. This
suggests that all these sources are powered by a very similar
(jet/disk) engine.  In this context it is quite interesting to note
that M81* and Sgr A* are in spiral galaxies, yet appear radio loud in
these diagrams. M81 also has recently shown double-peaked broad
emission lines, a feature usually seen only in radio-loud galaxies
(Bower et al. 1996).

Like in quasars and Sgr A*, one has to choose the most efficient,
equipartition jet model to explain M81*, with high internal energy and
a jet power that equals the disk luminosity. This leaves one with no
free parameters for M81* other than the jet inclination angle which
according to the current model is around $i\sim 30-40^\circ$. Some
caution is, however, neccessary, as the PA of the VLBI component seems
to change with frequency, something we have ignored here. It is not
clear whether this indicates bending, helical motion, or an extrinsic
effect and thus is difficult to interpret. Some of these effects could
change the results slightly -- but not to an order of magnitude.

In this {\it Letter} we have considered for the first time the effects
of the longitudinal pressure gradient in the BK79 and FB95 jet model.
Without any sophisticated mechanisms, this gradient alone will already
lead to a moderate acceleration of the jet to bulk Lorentz factors of
2-3. Consequently, the Lorentz factor will increase towards lower
frequencies and for a fixed viewing angle outside the boosting cone
the flux will become Doppler-dimmed with respect to higher
frequencies.  This can provide a natural explanation for the inverted
spectrum seen in compact radio cores in general and for the size/frequency
relation of M81*. Such a mildly accelerating jet may also be of
interest for quasar and BL Lac radio cores or galactic jet
sources. The biggest advantage, however, is that the jet velocity,
which was previously a free parameter, now is fixed by a simple
physical model.

Another interesting point of the presented model is the usage of an
initial, quasi monoenergetic electron distribution for which we find
$\gamma_{\rm e}\sim200$. This not only reduces another free parameter
(the electron distribution index) but also can explain the submm-IR
cut-offs seen in Sgr A* and M81*. If the jet in M81* does not have
strong shocks to re-accelerate those electrons into the ususal
power-law, the jet would have size-dependent high-frequency cut-offs
and thus explain the apparent absence of extended VLBI components in
M81*.

\acknowledgements This research was supported by NASA under grants
NAGW-3268 and NAG8-1027. I thank an anonymous referee for helpful
comments, and P. Biermann, W. Duschl, H. Lesch, P. Mezger and A. Wilson
for numerous discussions.


\figcaption[]{Parameter space for the jet/disk symbiosis model for
M81* ($D=3.25$ Mpc, $L_{\rm disk}=10^{41.5}$ erg/sec); the vertical
axis gives the size, the horizontal axis the flux of the radio
core. Values predicted by the model are calculated for various
inclination angles (circles) and proton/electron ratios $\mu_{\rm
p/e}$. The corresponding, predicted spectral and size indices depend
almost exclusively on $i$ and are $\alpha=(0.03, 0.09, 0.15, 0.19,
0.21, 0.23, 0.25, 0.26, 0.27)$ and $m=(-0.98, -0.95, -0.92, -0.9,
-0.89, -0.89, -0.88, -0.88, -0.88)$ for inclination angles $i=10^\circ
- 90^\circ$ in steps of $10^\circ$.  A model without longitudinal
pressure gradient would have predicted $\alpha=0$ and $m=-1$ for all
$i$. The black dots indicate how the parameter plane shifts if one
increases the jet power ($q_{\rm j/l}\cdot L_{\rm disk}$) by a factor
3.}


\begin{figure*}
\plotone{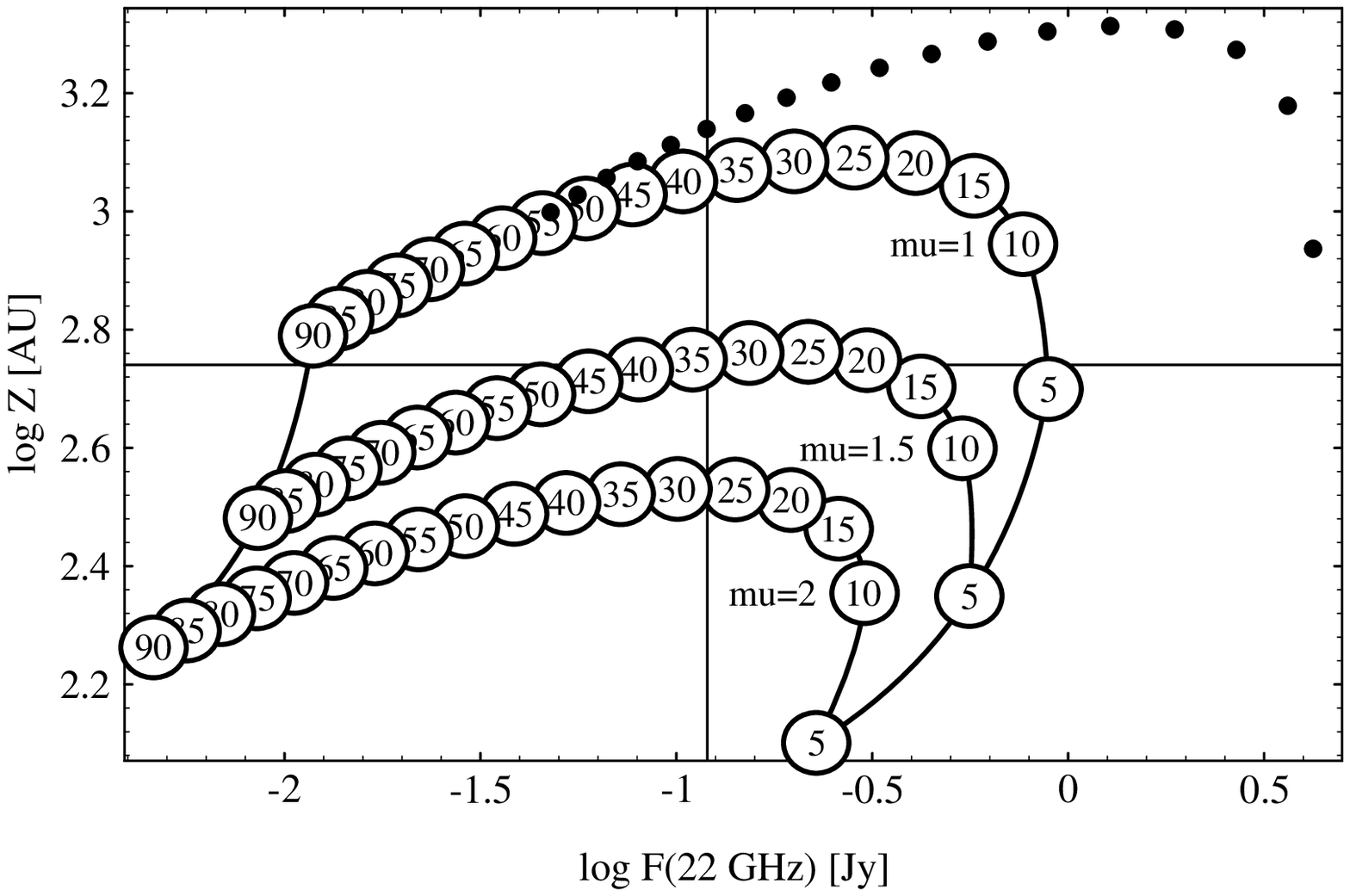}
\end{figure*}

\end{document}